\documentstyle[a4wide,epsf,11pt,titlepage]{article}

\newcounter{nref}
\newcommand{\bbib}{%
  \renewcommand{\refname}{\large\bf References}%
  \begin{thebibliography}{99}%
  \setcounter{enumiv}{\thenref}}
\newcommand{\ebib}{%
  \end{thebibliography}%
  \setcounter{nref}{\arabic{enumiv}}}
\newcommand{\head}[3]{%
  \setcounter{nref}{0}%
  \thispagestyle{empty}%
  \section*{\LARGE\bf #1}%
  \stepcounter{section}%
  \addcontentsline{toc}{section}{#1}%
  \large\itshape%
  #2\\\vspace{0.1pt}\\%
  #3%
  \normalsize\upshape%
  \bigskip}

\begin{document}

\head{Ionization freeze-out and barium problem in supernova 1987A}
     {V.P.\ Utrobin$^{1,2}$, N.N.\ Chugai$^3$}
     {$^1$ Institute for Theoretical and Experimental Physics,
           117259 Moscow, Russia \\
      $^2$ Max-Planck-Institut f\"ur Astrophysik,
           D-85741 Garching, Germany\\
      $^3$ Institute of Astronomy of Russian Academy of Sciences,
           109017 Moscow, Russia }

\subsection*{Abstract}

 We have shown that in the atmosphere of SN 1987A level populations of hydrogen are
    mostly controlled by an ionization freeze-out and ion-molecular processes up to
    $\sim$ 40 days.
 The ionization freeze-out effects are important for normal SNe II-P as well.
 The time dependent effects and the mutual neutralization between H$^{-}$ and
    H$^{+}$ may result in a non-monotonic radial dependence of the Sobolev optical
    depth for H$\alpha$ and then in a blue emission satellite of H$\alpha$ observed
    in SN 1987A spectra at Bochum event phase.
 The relative abundance of molecular hydrogen n(H$_{2}$)/n(H) $\sim$ 10$^{-4}$--10$^{-3}$
    is high enough for the photospheric phase of SN 1987A.
 We emphasize that the poor knowledge of the far UV radiation field  in the envelope
    of SN 1987A still forbids a truly reliable estimate of the Ba abundance but
    well-constructed and tested models may improve this situation.

\subsection{Introduction}

 As time goes on an impression strengthens that almost everything is well
    understood with respect to supernova (SN) 1987A in the Large Magellanic Cloud
    (LMC).
 But it is not the case.
 Let us consider merely two neighbor lines: H$\alpha$ and Ba II 6142 \AA.
 At early times (during 20--100 days) the H$\alpha$ line exhibits a striking fine
    structure called "Bochum event" \cite{viunic.1}.
 This phenomenon was originally described as two additional emission-like
    features, a red emission satellite (RES) and a blue emission satellite (BES).
 Such an H$\alpha$ profile was supposed to result from a superposition of an
    asymmetric and a spherically symmetric component \cite{viunic.2}.

 The asymmetric component responsible for the RES detail consisted of the bright
    core with the radial velocity of $+4000$ km\,~s$^{-1}$ and the extended halo
    of a lower brightness.
 A transversal velocity of the core of 2400 km\,~s$^{-1}$, estimated with the effect
    of the occultation of the asymmetric component by the photosphere, together with
    the radial velocity gave the absolute velocity of 4700 km\,~s$^{-1}$.
 It was evident that the core of the asymmetric component coincided with the $^{56}$Ni
    clump, which should have the same absolute velocity of 4700 km\,~s$^{-1}$.
 The amount of $^{56}$Ni in the fast clump was estimated as $\sim 10^{-3} M_\odot$.
 The symmetric component was adapted to match the BES feature of H$\alpha$ and was
    characterized by the non-monotonic radial dependence of the Sobolev optical
    depth which had no physical explanation.

 The unusually strong Ba II 6142 \AA\ line in early spectra of SN~1987A is
    a distinctive feature of this supernova \cite{viunic.3}, which still remains
    a subject of great concern after first studies \cite{viunic.4, viunic.5}
    claimed the large (up to a factor of 20) Ba overabundance derived from the
    line strength.
 The problem is that the $s$-process nucleosynthesis in massive stars evolving to
    the presupernova star of SN~1987A \cite{viunic.6} is not able to yield the Ba
    overabundance in the hydrogen envelope by more than a factor of 1.4 assuming
    that a total ejecta mass is 15 $M_{\odot}$, a mass-cut is at 2~$M_{\odot}$
    \cite{viunic.7}, and the synthesized Ba is completely mixed over the ejecta.

 More naturally is to consider the large strength of Ba II 6142 \AA\ as the outcome
    of specific conditions in the SN~1987A atmosphere.
 It was recognized that if Ba in the envelope of SN~1987A was mostly in Ba II ion
    form then the Ba II 6142 \AA\ line might be reproduced with the Ba abundance
    typical for the LMC \cite{viunic.8}.
 The absence of strong Ba lines in normal SNe~II-P was explained then by the fact that
    Ba in the atmosphere of these supernovae was mostly in Ba III form.
 An atmosphere model based on solving the radiation transfer equation with the Monte
    Carlo technique was able to account for the low strength of Ba II lines in normal
    SNe~II-P for the solar abundance, but failed to produce strong Ba II lines in
    SN~1987A for the LMC abundance \cite{viunic.9}.
 It was acknowledged that this atmosphere model of SN~1987A used for Ba II lines
    simulation was presumably not adequate enough because of a poor agreement in
    hydrogen Balmer line intensities.

 Generally, until recently the modelling of hydrogen lines in SN~1987A at the
    photospheric epoch has remained a challenging problem for spectrum
    synthesis (e.g.\ \cite{viunic.10}).
 Situation has changed recently, when it has become clear that the major drawback of
    standard atmosphere models for SNe~II-P was an assumption of the statistical
    equilibrium.
 It has been shown in the frame of time dependent chemical kinetics that an ionization
    freeze-out plays a crucial role in the ionization and excitation of hydrogen in
    the atmosphere of SNe~II-P at the photospheric epoch \cite{viunic.11, viunic.12}.
 Moreover, in the atmosphere of SN~1987A ion-molecular processes might become
    essential in producing ionization of hydrogen.
 In view of these results a motivation of the work is to take time dependent effects
    into account, to construct an adequate model, to succeed in reproducing H$\alpha$
    profile, and then to study the Ba II 6142 \AA\ line in SN~1987A at the photospheric
    epoch.

\begin{figure}[t]
  \centerline{\epsfxsize=0.45\textwidth\epsffile{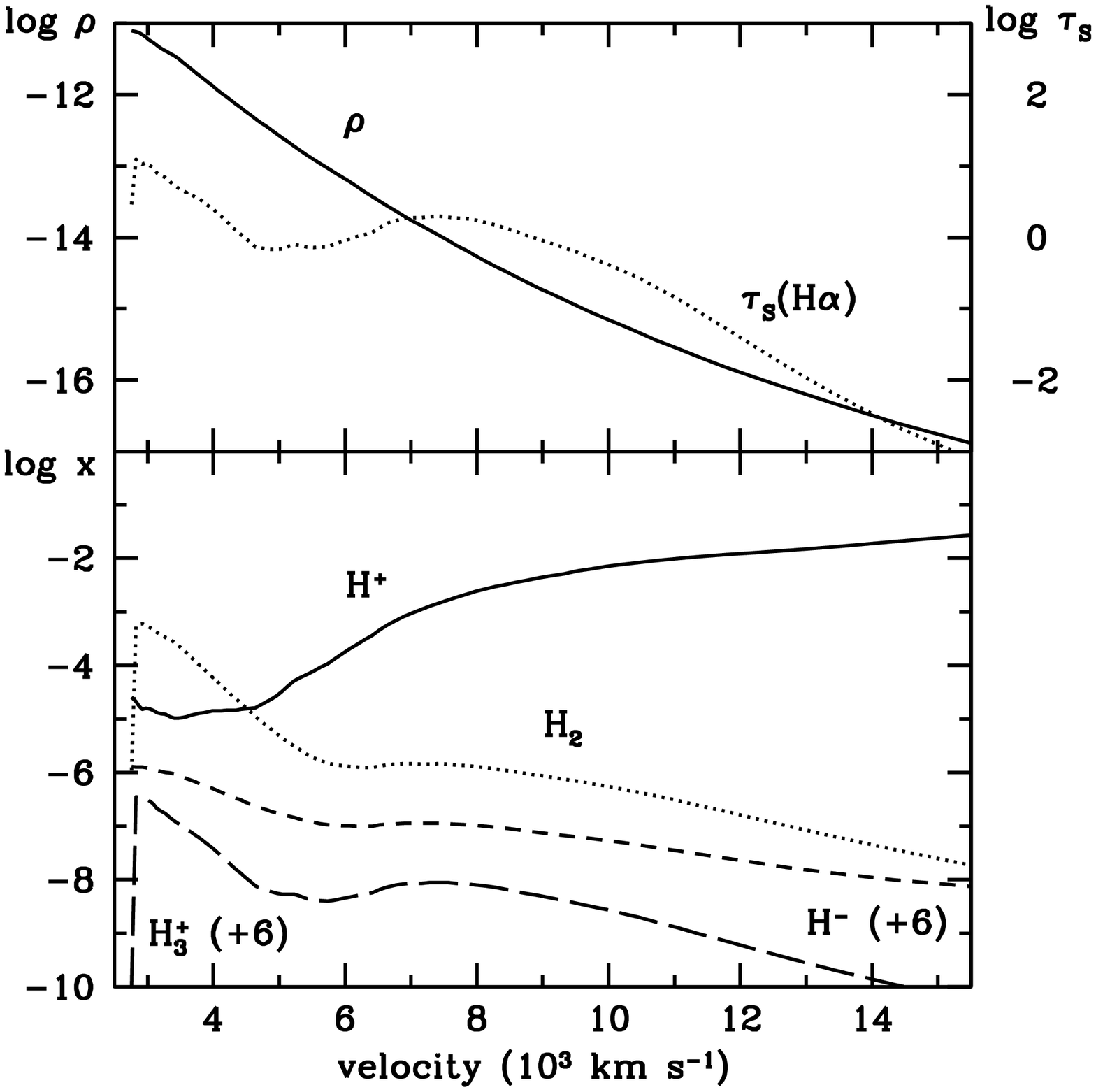}
              \epsfxsize=0.45\textwidth\epsffile{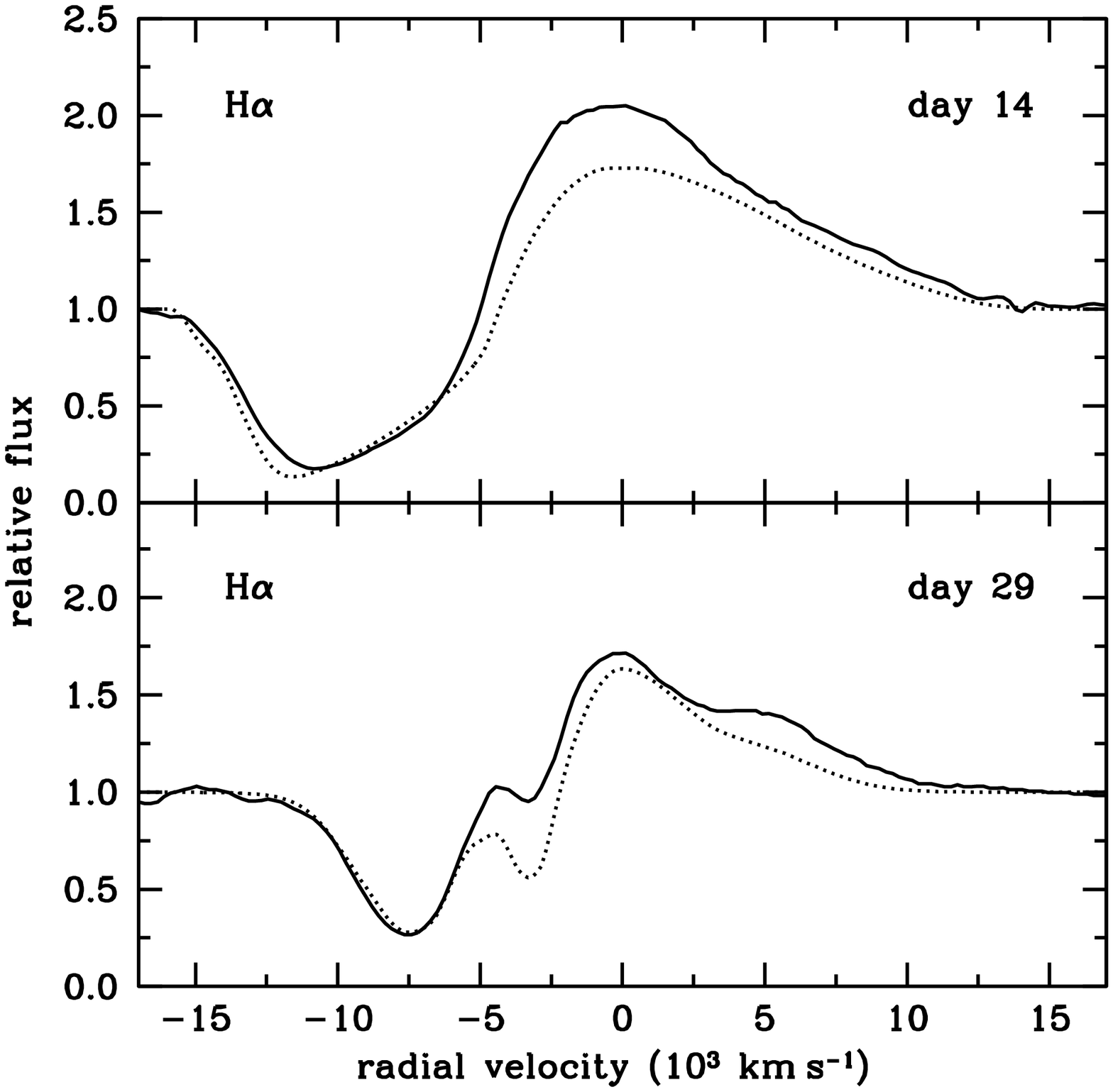}}
  \caption{Upper left panel: Density (solid line) and the Sobolev optical depth
           of H$\alpha$ (dotted line) as a function of velocity for day 29.
           Lower left panel: Velocity dependence of the following fractional
           abundances for day 29: x(H$^{+}$) (solid line), x(H$_{2}$) (dotted line),
           x(H$^{-}$) (short dashed line), and x(H$_{3}^{+}$) (long dashed line).
           Dashed lines are shifted by a $+6$ order of magnitude.
           Right panel: Observed (solid line) and calculated (dotted line)
           H$\alpha$ profile for days 14 and 29.}
  \label{viunic.fig1}
\end{figure}

\subsection*{Model and Input Physics}

 A full model should be based on the hydrodynamic model and time-dependent
    radiation transport, and should include self-consistent electron temperature
    evolution and full reaction network.

 Our present model is based on the hydrodynamic model \cite{viunic.13} which has an
    ejecta of mass 15$M_{\odot}$ and a kinetic energy of 1.44$\times$10$^{51}$ erg.
 The radiation field is treated in the approximation of clear-cut photosphere and
    above atmosphere.
 For t$<$1.8 days the radiation field in continuum is described by the photospheric
    radius and effective temperature, taken from the hydrodynamic model, and then by
    those observed \cite{viunic.14} and by the approximated UV and optical observations
    \cite{viunic.15}.
 The line radiation transfer is treated in the modified Sobolev approximation
    \cite{viunic.16, viunic.17} as a purely local process.
 Instead of solving the energy equation we use two regimes for the electron
    temperature: radiative equilibrium and adiabatic approximation.

 The following elements and molecules are calculated in non-LTE chemical kinetics:
    H, He, C, N, O, Ne, Na, Mg, Si, S, Ar, Ca, Fe, Ba, H$^{-}$, H$_{2}$, H$_{2}^{+}$,
    and H$_{3}^{+}$.
 All elements but H are treated with the three ionization stages.
 The level populations of H and Ba II are calculated for 15 and 17 levels,
    respectively, and the rest of atoms and ions are assumed to consist of
    the ground state and continuum.
 The reaction network involves all bound-bound and bound-free, radiative and
    collisional processes for atoms and ions, and 7 radiative and 37 collisional
    processes for molecules.
 To reproduce the chemical composition typical to the LMC situation \cite{viunic.18}
    all metal abundances are scaled to 1/2.88 their solar values.

\begin{figure}[t]
  \centerline{\epsfxsize=0.45\textwidth\epsffile{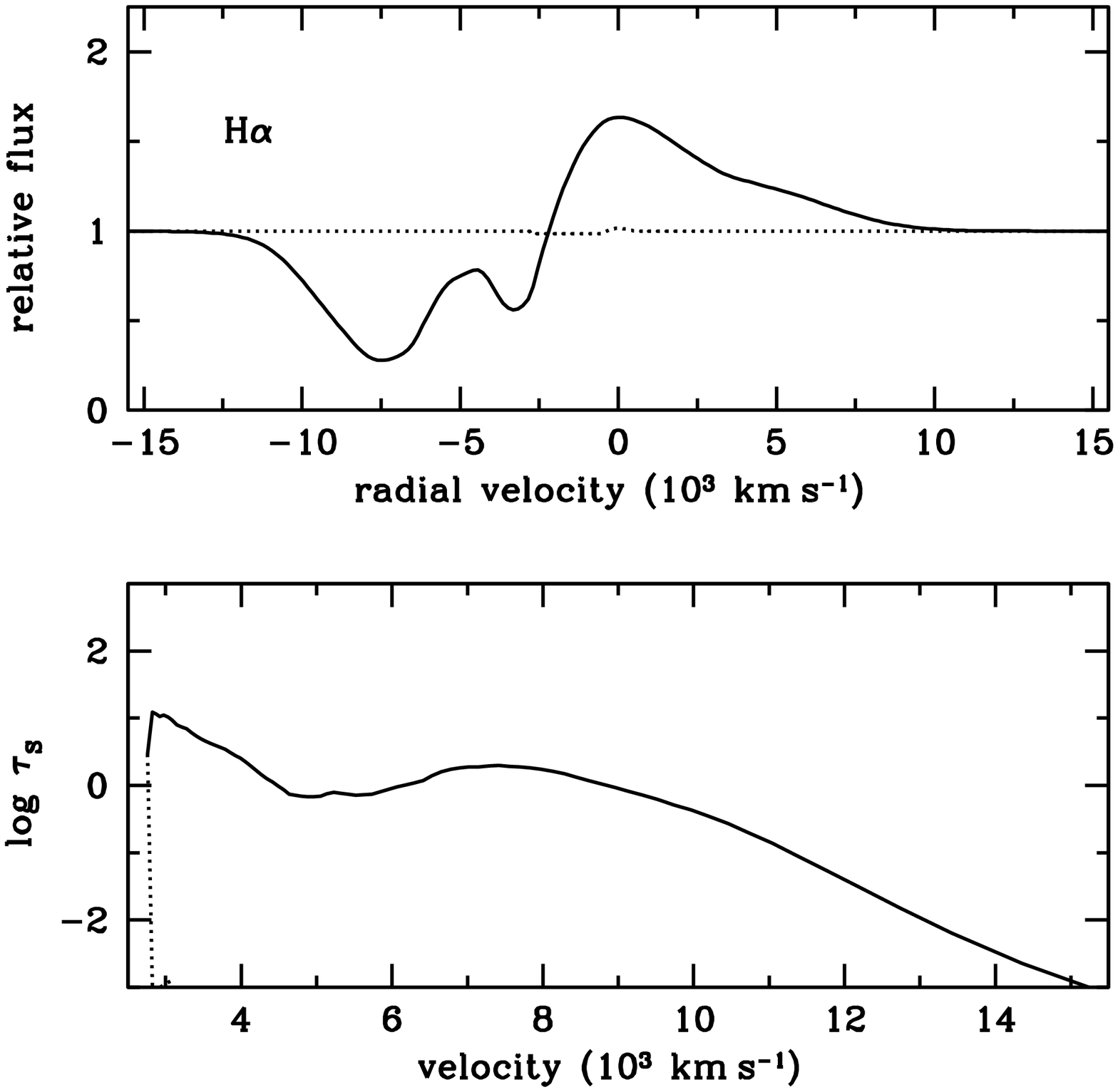}
              \epsfxsize=0.45\textwidth\epsffile{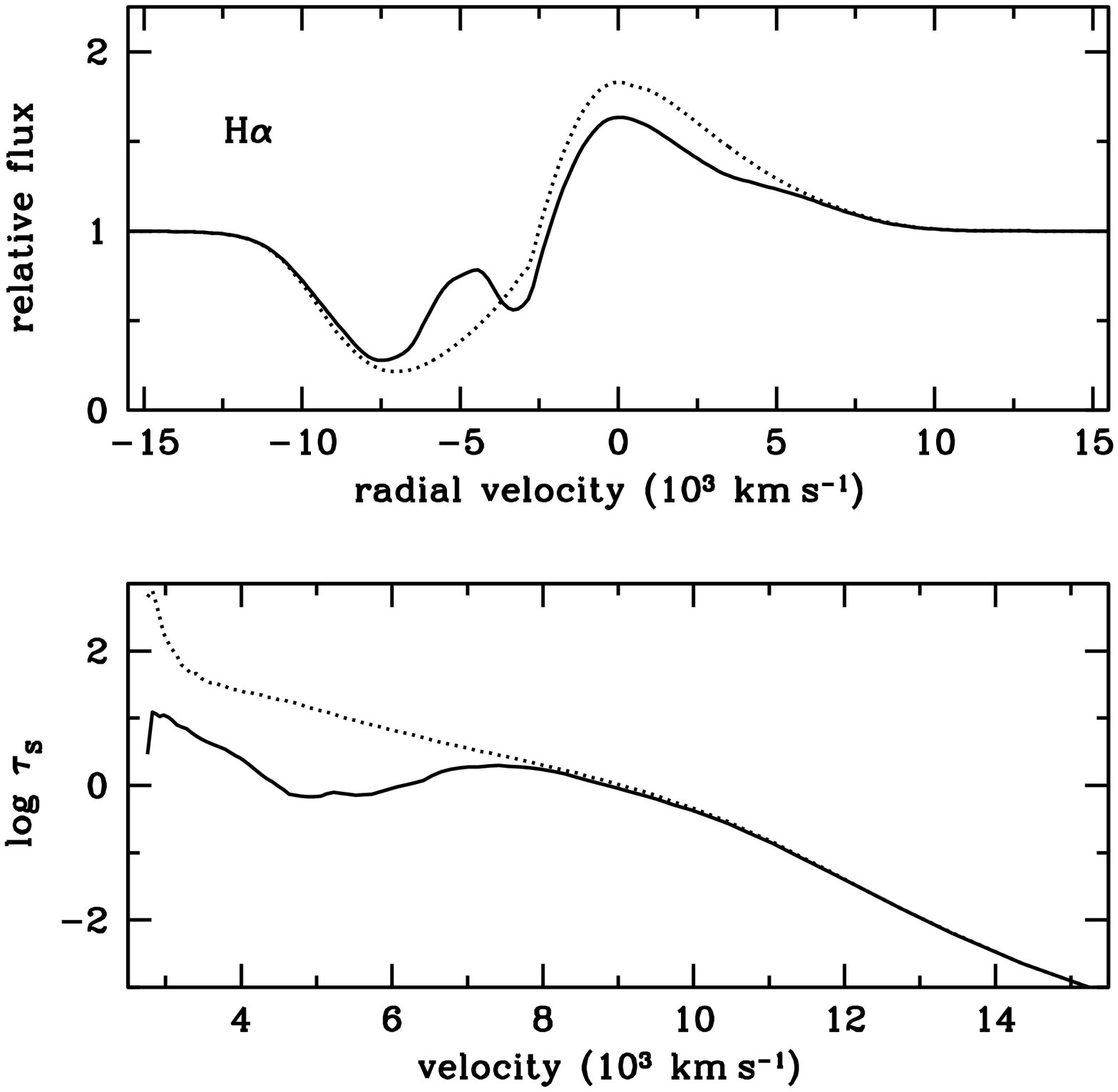}}
  \caption{Calculated H$\alpha$ profile and the Sobolev optical depth as a function
           of velocity for day 29. Influence of non-stationarity (left panel) and
           molecules (right panel) is shown by dotted line.}
  \label{viunic.fig2}
\end{figure}

\subsection*{Results}

 We have developed a time dependent chemistry model for supernova envelope at
    photospheric phase and investigated spectra of SN 1987A.
 On day 29 Fig.~\ref{viunic.fig1} shows the specific non-monotonic radial
    dependence of the Sobolev optical depth of H$\alpha$ over envelope with
    minimum near the photosphere.
 Just such a distribution we need to reproduce the BES feature of H$\alpha$ profile.
 Note that the fractional abundance of molecular hydrogen of the order of 10$^{-4}$
    is high enough for the photospheric phase of SN 1987A.
 The fit between the observed and calculated H$\alpha$ profiles on days 14 and 29
    is fairly good in Fig.~\ref{viunic.fig1} indicating that we have constructed
    a correct model.

 The influence of time dependent effects on the H$\alpha$ profile is shown in
    Fig.~\ref{viunic.fig2}.
 It is clear that the ionization freeze-out plays a key role in producing the
    ionization and excitation of hydrogen for times nearly up to day 40 when the
    non-thermal ionization and excitation resulting from radioactive $^{56}$Ni decays
    become essential \cite{viunic.13}.
 In addition, from Fig.~\ref{viunic.fig2} it is evident that molecules are
    mainly responsible for the formation of the minimum of the Sobolev optical depth
    above the photosphere and, as a consequence, for the BES feature at Bochum
    event phase.
 A leading reaction in the neutralization of ionized hydrogen is mutual
    neutralization between H$^{-}$ and H$^{+}$: H$^{-}$ + H$^{+}$ $\rightarrow$ 2 H.

\begin{figure}[t]
  \centerline{\epsfxsize=0.45\textwidth\epsffile{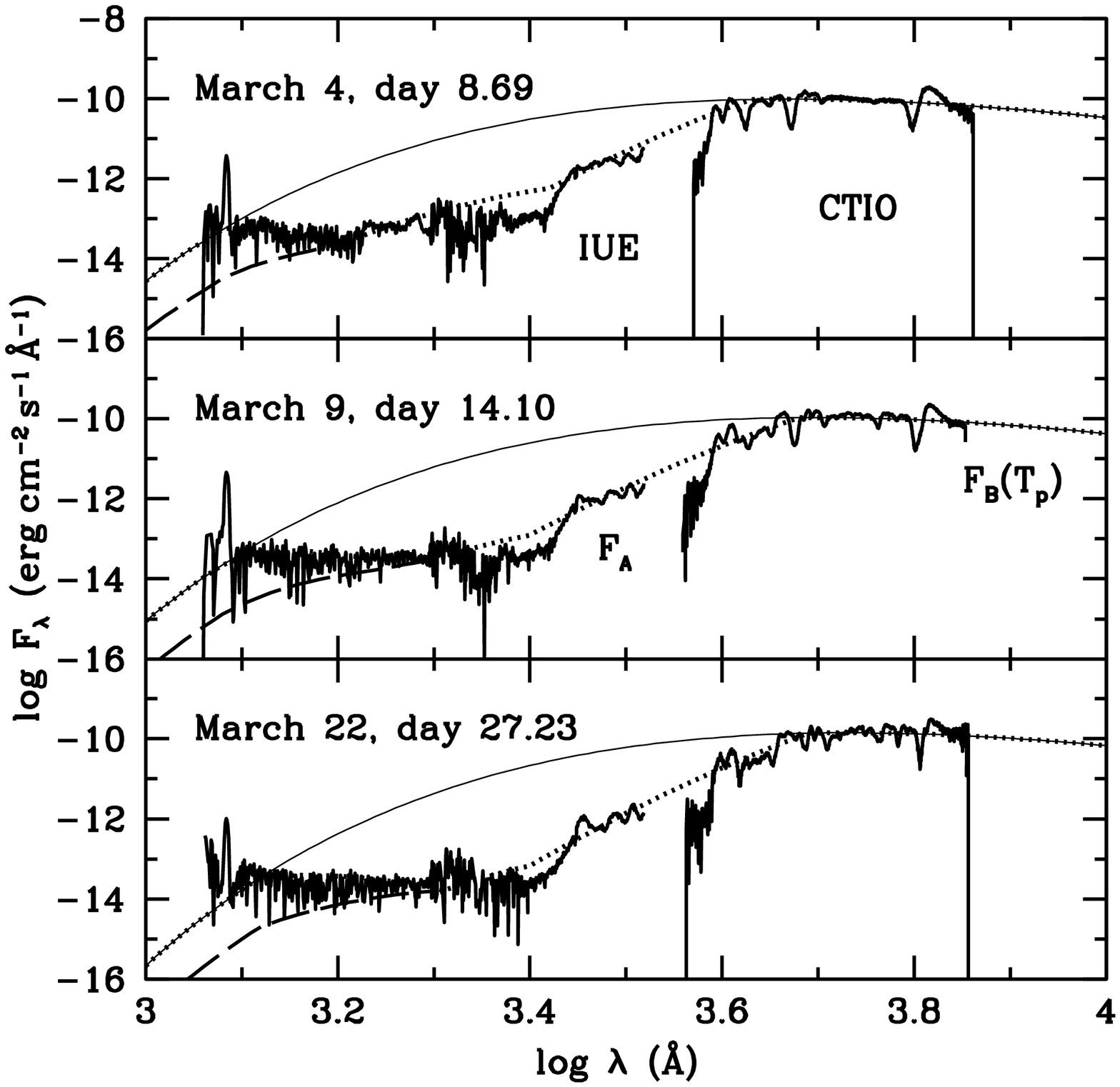}
              \epsfxsize=0.45\textwidth\epsffile{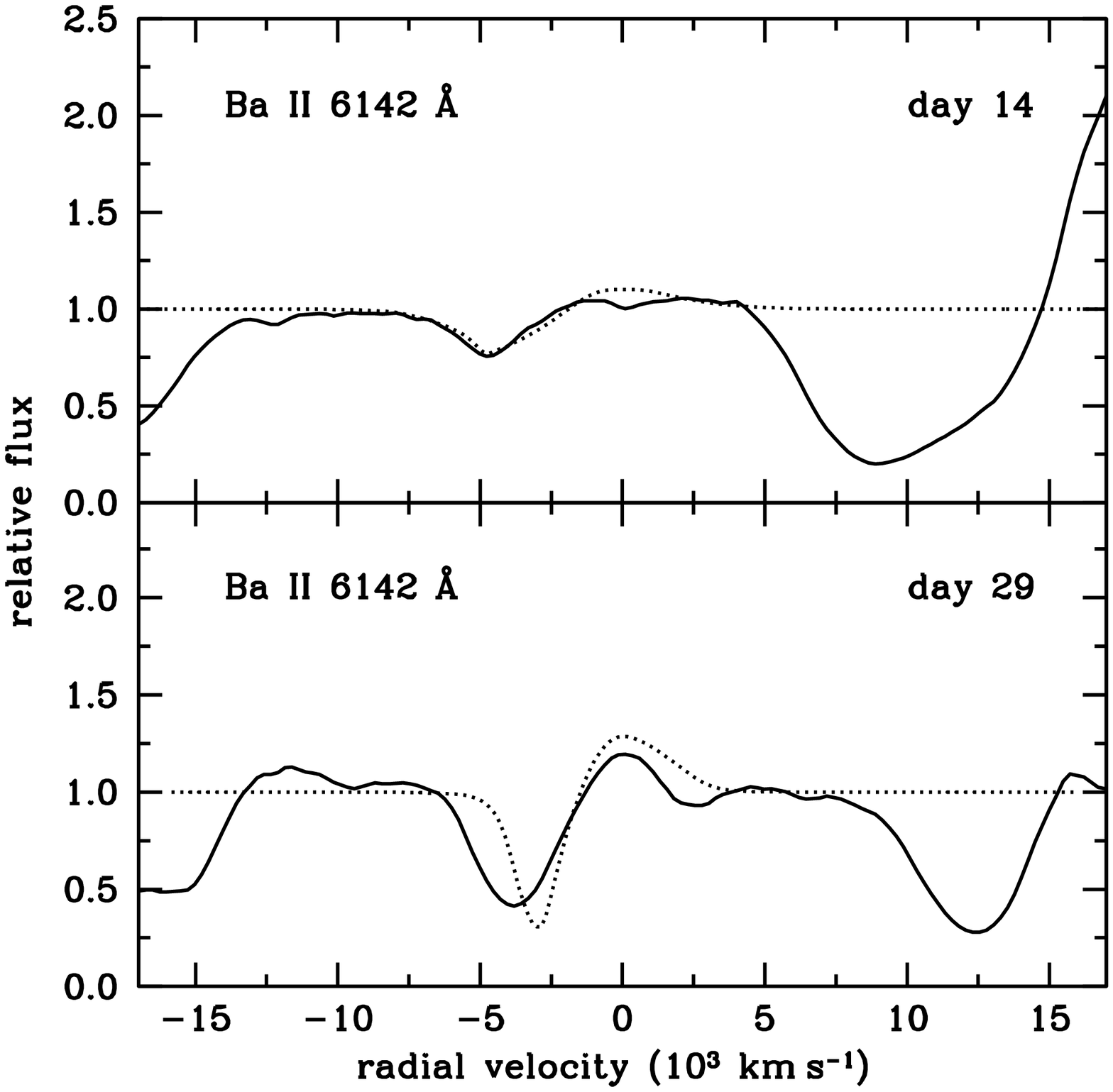}}
  \caption{Left panel: The combined UV (IUE) and optical (CTIO) spectra of
           SN 1987A \cite{viunic.15} (thick solid line), the black-body flux
           at the effective temperature (thin solid line), the approximated
           emergent flux (dotted line), and the approximated emergent flux
           reduced by a factor of $\approx$ 17 at log $\lambda < 3.2-3.3$
           (long dashed line) for days 8, 14, and 27.
           Right panel: Observed (solid line) and calculated (dotted line)
           Ba II 6142 \AA\ line for days 14 and 29.}
  \label{viunic.fig3}
\end{figure}

 Now when we have got a confidence in reproducing the observed H$\alpha$ profile
    we can study the barium line.
 In Fig.~\ref{viunic.fig3} there is a good fit between the observed and calculated
    Ba II 6142 \AA\ line but the approximated UV and optical observations result
    in the Ba overabundance ratio $\approx$ 12 for the time-dependent solution and
    $\approx$ 16 for the steady one.
 The point is that the fractional abundance of Ba II in the supernova envelope
    turns out too low to produce the strong barium line with the low Ba
    overabundance ratio.
 We have analyzed this situation and found out that, it is evident, the fractional
    abundance of Ba II is very sensitive to the radiation flux at photon energies
    responsible for Ba II ionization from ground state and excited levels.

\subsection*{Discussion}

 To evaluate the influence of the radiation flux on the Ba II fractional
    abundance we calculate a toy model with the Ba abundance ratio of 1.4 as
    predicted from the $s$-process nucleosynthesis.
 In this model the calculated Ba II 6142 \AA\ line matches the observed one
    as shown in Fig.~\ref{viunic.fig3} assuming a reduction of the radiation
    flux by a factor of $\approx$ 17 in the far UV region.

 So, the far UV radiation field plays a vital role in estimating Ba abundance
    in SN 1987A.
 A crucial question is whether the observed flux is an intrinsic flux of SN 1987A
    or not.
 Taking a number of uncertainties in measuring the far UV flux of SN 1987A into
    account it is reasonable to assume that the observed flux is presumably not related to
    the radiation field in the region where the barium line forms.
 Thus we have at least two alternatives: pessimistic and optimistic.
 If we consider the observed flux as responsible for the Ba II ionization in
    supernova envelope we should accept the Ba abundance as large as 12 or so.
 And there is nothing more to do.
 This is the pessimistic alternative.
 In opposite case we can solve the radiation transfer equation in far UV region,
    check the solution in some way, and estimate Ba abundance in SN 1987A.
 It is the optimistic alternative.

\subsection*{Conclusions}

 We have developed a time dependent chemistry model for supernova envelope at
    photospheric phase that, in connection with hydrodynamic models, provides
    a powerful tool to investigate supernova spectra.

 We have shown that level populations of hydrogen are mostly controlled by an
    ionization freeze-out and ion-molecular processes up to $\sim$ 40 days in
    the atmosphere of SN 1987A. The ionization freeze-out effects are important
    for normal SNe II-P as well.

 The time dependent effects and mutual neutralization between H$^{-}$ and
    H$^{+}$ may result in a non-monotonic radial dependence of the Sobolev optical
    depth for H$\alpha$ and then in a blue emission satellite of H$\alpha$ observed
    in SN 1987A spectra at Bochum event phase.

 It should be noted that the relative abundance of molecular hydrogen
    n(H$_{2}$)/n(H) $\sim$ 10$^{-4}$--10$^{-3}$ is high enough for the photospheric
    phase of SN 1987A.

 We emphasize that the poor knowledge of the far UV radiation field in the envelope
    of SN 1987A still forbids a truly reliable estimate of the Ba abundance but
    well-constructed and tested models may improve this situation.

\subsection*{Acknowledgements}

 V.P.U. is grateful to W. Hillebrandt and E. M\"{u}ller for hospitality, for
    giving an opportunity to participate in the Eleventh Workshop on Nuclear
    Astrophysics, and for the financial support of his participation at this
    workshop and staying at the MPA, and also would like to thank Keith Butler
    for providing him with atomic data on the once-ionized barium.
 The work in Russia is partially supported by the RFBR (project 01-02-16295).

\end{document}